\documentclass[twocolumn]{autart}    
\usepackage{amssymb,amsmath,textcomp}
\usepackage{color}
\usepackage{graphicx}          
 \usepackage{algorithm} 
\usepackage{multirow}
\begin{document}

\begin{frontmatter}

\title{Risk Sensitive Filtering with Randomly Delayed Measurements\thanksref{footnoteinfo}} 

\thanks[footnoteinfo]{This paper was not presented at any IFAC 
meeting. Corresponding author R.~K.~Tiwari.}

\author[India]{Ranjeet Kumar Tiwari$^\ast$}\ead{ranjeet.pee16@iitp.ac.in},    
\author[India]{Shovan Bhaumik}\ead{shovan.bhaumik@iitp.ac.in}               

\address[India]{Department of Electrical Engg., Indian Institute of Technology Patna, Patna, India}  
\date{}
          

\begin{abstract}                          
Conventional Bayesian estimation requires an accurate stochastic model of a system. However, this requirement is not always met in many practical cases where the system is not completely known or may differ from the assumed model. For such a system, we consider a scenario where the measurements are transmitted  to a remote location using a common communication network and due to which, a delay is introduced while receiving the measurements. The delay that we consider here is random and one step maximum at a given time instant. For such a scenario, this paper develops a robust  estimator for a linear Gaussian system by minimizing the risk sensitive error criterion that is defined as an expectation of the accumulated exponential quadratic error. The criteria for the stability of the risk sensitive Kalman filter (RSKF) are derived and the results are used to study the stability of the developed filter. Further, it is assumed that the latency probability related to delay is not known and it is estimated by maximizing the likelihood function. Simulation results suggest that the proposed filter shows acceptable performance under the nominal conditions, and it performs better than the Kalman filter for randomly delayed measurements and the RSKF in presence of both the model uncertainty and random delays.
\end{abstract}
\end{frontmatter}

\section{Introduction}

An optimal solution which is popularly known as the Kalman filter (KF) is available to estimate the state of a linear Gaussian system \cite{anderson2012optimal}.
 However, it needs to be modified if the measurements are received with random delays \cite{ray1993state}.  Consequently, a considerable amount of research work is reported in literature for estimating the states of a linear Gaussian system with randomly delayed measurements \cite{sun2014modeling,ma2017distributed,ma2011optimal}. An unbiased minimum variance filtering solution is discussed in \cite{ray1993state} for a linear Gaussian system with one step random delay in measurements. For the system with bounded measurements delay and packet drops, a minimum variance filter with the state augmentation approach is proposed in \cite{sun2009linear}. The authors of \cite{he2013optimal} used the linear temporal coding technique to design an optimal linear estimator for a networked system with packet dropping events. The works in \cite{zhang2011linear} and \cite{sun2012optimal} dealt with the measurements which are randomly delayed but time stamped. The authors in \cite{zhang2011linear} has adopted a method of measurement reorganization for designing the optimal estimator, whereas \cite{sun2012optimal} has used the state augmentation method.  Note that none of the above works entertains the model uncertainty in its process dynamics.

Many a time in real practice, the system does not adhere to the assumed model, and the value of its model parameters deviates from the nominal. For such systems, several robust estimators based on different cost functions have been reported in literature \cite{haddad1987optimal,haddad1991mixed,xie1994robust,dey1995topics,bhaumik2009risk,sadhu2009particle}. A study on the state estimation of a linear Gaussian system with exponential performance criteria is carried out in \cite{jacobson1973optimal} to tackle a deterministic model uncertainty. The dynamic programming method with the same performance criterion is proposed in \cite{speyer1992optimal} to provide the required robustness. The authors in \cite{dey1997risk} and \cite{boel2002robustness} considered the risk sensitive filtering solutions to handle such uncertainty in system. However, none of these works has addressed the simultaneous presence of random delays in measurements and the uncertainty in system model while designing a state estimator.

 The model uncertainty that we are entertaining in this work is a part of the plant dynamics and implies that the values of one or more process parameters deviate from the nominal ones and are not known correctly. These uncertainties in their values are deterministic, arbitrary, and unknown. This should not be confused with the process noise, which is a random sequence with known distribution. The measurements are randomly delayed with one step maximum at a given time instant, and it is modeled with help of the Bernoulli random variables. This work considers the exponential of the squared estimation errors, both the past and the present, as a cost function and presents a framework to obtain a general solution.  Subsequently, the cost function is minimized for a linear Gaussian system with above descriptions and we receive a closed form recursive solution.
 
  Further, it is assumed that the latency probability of the random delays is not known, and we propose a method to estimate it based on the joint density of the received measurements. This paper also establishes the stability of the RSKF using the uniformly complete controllability and observability condition. This result is further utilized to analyze the stability of the proposed method and presented in the form a conjecture. The developed estimator converges to the RSKF in absence of random delays in measurements and becomes the KF in absence of both the random delays and the model uncertainty.

The proposed filter is applied to two linear estimation problems and its performance is compared with that of the RSKF \cite{boel2002robustness} and the Kalman filter for randomly delayed measurements (KF-RD) \cite{ray1993state}. The simulation results suggest that the proposed filter shows acceptable performance (comparable to the Kalman filter) under the nominal conditions and performs better than the KF-RD and the RSKF when the system deviates from the nominal attributions.

The specific contributions of this paper over the existing works can be summarized as follows:
\begin{enumerate}
	\item[(i)] It derives a closed form solution for a linear Gaussian system in presence of the model uncertainty in process dynamics and single step random delays in the measurements.
	\item[(ii)] The latency probability in measurement model is also estimated by maximizing the likelihood of the received measurements.
	\item[(iii)] Further, it establishes the stability criteria of the RSKF and then utilize the result to justify the stability of the proposed filter. 
\end{enumerate}

The rest of the paper is organized as follows. A recursive Bayesian framework using the information state and the risk sensitive error criteria for a system with the model uncertainty and random delays in measurements is presented in Section~2. Section~3 uses the framework developed in the previous section and derives the recursive algorithm of the proposed filter.  Section~4 discusses the stability of both the RSKF and the proposed filter. The estimation of latency parameter for randomly delayed measurements is presented in Section~5. In Section~6, the simulations results are listed. Finally, the paper ends with a brief conclusion.


\section{Risk sensitive filtering with randomly delayed measurements}
Let us consider a discrete time nonlinear system, modeled with the following process and measurement equation:
\begin{equation}\label{gen_sys}
	x_k=f_{k-1}(x_{k-1})+w_{k-1},
\end{equation}
\begin{equation}\label{nd_meas}
	z_k=h_k(x_k)+v_k,
\end{equation} 
where the state $x_k\in \Re^{n_x}$, the measurement $z_k \in \Re^{n_z}$, $f_k:\Re^{n_x}\rightarrow \Re^{n_x}$, and $h_k:\Re^{n_x} \rightarrow \Re^{n_z}$. The process noise $w_{k-1}$ and the measurement noise $v_k$ are independent and identically distributed (i.i.d.) random processes with arbitrary but known probability density function (pdf). We consider that the actual system differs from the assumed model and follows the following process dynamics:
\begin{equation}\label{actual_process}
	x_k=f_{k-1}(x_{k-1})+\Delta f_{k-1}(x_{k-1})+w_{k-1},
\end{equation} where $\Delta f_{k-1}(x_{k-1})$ represents an arbitrary, deterministic and unknown process modeling error.

The measurements are assumed to be transmitted over a common communication channel to a remotely located estimation center and owing to the limited bandwidth of the channel, a delay is introduced during the transmission. We assume that the delays are random in nature and the maximum extent of it is one step. The received measurement, $y_k$, can be modeled as \cite{ray1993state}
\begin{equation}\label{meas}
	y_k=(1-\beta_k)z_k+\beta_kz_{k-1},
\end{equation} where $\beta_k$ are the i.i.d. random sequences that follow the Bernoulli distribution with $\mathbb{E}[\beta_k]=\alpha_k$, and $\mathbb{E}[\beta_i\beta_j^\top]=\mathbb{E}[\beta_i]\mathbb{E}[\beta_j],\ \forall i\neq j$.

\subsection{Approach}
Our objective is to find an optimal posterior estimate $\hat{x}_{k|k}^\ast$ recursively from the remotely received measurements $y_{1:k}$ for the underlying system described in (\ref{gen_sys}-\ref{meas}). We  consider a risk sensitive cost criterion given by 
\begin{equation}\label{cost}
	\begin{split}
	J_k(\hat{x}_{k|k}|y_{1:k})=&\mathbb{E}_\beta \mathbb{E}_x\Big[\exp\Big(\sum_{i=1}^{k-1}\mu_{1,i}e_{i|i}^{\ast\top} e_{i|i}^{\ast}\\& + \mu_{2,k} e_{k|k}^\top e_{k|k}\Big)\Big],
	\end{split}
\end{equation} 
where the posterior estimation error is $e_{i|i}^\ast=(x_i - \hat{x}_{i|i}^\ast)$.  $\hat{x}_{i|i}$ denotes the posterior estimated state at time step $i$.  $\mu_{1,i}\geq 0$ and $\mu_{2,i}>0$ are two time varying risk parameters used for scaling the past errors and the present error, respectively. $\mathbb{E}_x$ and $\mathbb{E}_\beta$ denote the expectations over the posterior density of $x_k$, and  the statistics of $\beta_k$, respectively. Hereafter, for simplicity $\mathbb{E}_\beta \mathbb{E}_x[.]$ will be written as $\mathbb{E}[.]$. We seek the state estimate at each time step which minimizes the cost function expressed in (\ref{cost}), that is
\begin{equation}\label{min_fun}
	{\hat{x}_{k|k}^\ast}=\arg \ \min\limits_{{\hat{x}_{k|k}}} J_k(\hat{x}_{k|k}|y_{1:k}).
\end{equation}

\textbf{Remark~1.}  Note that the posterior density of the state is $p(x_k|y_{1:k}) \propto p(y_k|x_k,x_{k-1})p(x_k|x_{k-1})$, where the densities, $p(y_k|x_k,x_{k-1})$ and $p(x_k|x_{k-1})$, are constructed using the process and measurement equations, and the pdf of $v_k$ and $w_{k-1}$.

\subsection{General framework}
In this subsection, we work out a general framework for the solution of (\ref{min_fun}) by using the two step Bayesian framework for the risk sensitive error criterion.  The joint posterior density with the delayed measurements, $y_{1:k}$, can be written as:
\begin{equation}\label{pre_simp_pos_den}
	\begin{split}
		p(x_{0:k}|y_{1:k})&=\dfrac{p(x_{0:k},y_{1:k})}{p(y_{1:k})}\\
		&=\dfrac{p(y_k|x_{0:k},y_{1:k-1})p(x_{0:k},y_{1:k-1})}{p(y_k|y_{1:k-1})p(y_{1:k-1})}\\
		&=\dfrac{p(y_k|x_{0:k},y_{1:k-1})p(x_{0:k}|y_{1:k-1})p(y_{1:k-1})}{p(y_k|y_{1:k-1})p(y_{1:k-1})}\\
		&=\gamma_kp(y_k|x_{0:k},y_{1:k-1})p(x_{0:k}|y_{1:k-1}),	
	\end{split}
\end{equation}
where $\gamma_k=1/ p(y_k|y_{1:k-1})$, and $p(y_k|y_{1:k-1})=\int p(y_k|x_{0:k},y_{1:k-1})p(x_{0:k}|y_{1:k-1}) dx_{0:k}$ is a normalizing constant. From (\ref{nd_meas}) and (\ref{meas}), it is clear that the current measurement $y_k$ is correlated with the current state $x_k$ as well as previous state $x_{k-1}$. Thus, assuming $y_k$, conditioned on $x_k$ and $x_{k-1}$, is independent of the previous measurements, $y_{1:k-1}$, and the states, $x_{0:k-1}$, we can write $p(y_k|x_{0:k},y_{1:k-1})=p(y_k|x_k,x_{k-1})$.  Assuming $\hat{x}_{0|0}^\ast, \cdots, \hat{x}_{k-1|k-1}^\ast$ are already known at the time step $k$ in (\ref{pre_simp_pos_den}),  we write the marginal density of state as
\begin{equation}\label{inter_simp_pos_den}
	p(x_k|y_{1:k})=\gamma_k p(y_k|x_k,x_{k-1})p(x_k|y_{1:k-1}).
\end{equation}
Again, using the Chapman-Kolmogorov integral for $p(x_k|y_{1:k-1})$, Eq. (\ref{inter_simp_pos_den}) can be rewritten as
\begin{equation}\label{simp_pos_den}
	\begin{split}
		p(x_k|y_{1:k})=&\gamma_k p(y_k|x_k,x_{k-1})\int p(x_k|x_{k-1})\\ & \qquad \times p(x_{k-1}|y_{1:k-1})dx_{k-1}.
	\end{split}
\end{equation} 
\textit{Information state}:
Consider a set of information, $\mathcal{I}_k=\{y_{1:k}, {e_{1|1}, \cdots, e_{k-1|k-1}\}}$ \cite{bar2004estimation}, is available at any time step $k$. The information state is defined as
\begin{equation}\label{info}
	\sigma_k\triangleq p(x_k|I_k)=\exp\Big(\sum_{i=0}^{k-1}\mu_{1,i}e_{i|i}^\top e_{i|i}\Big)p(x_k|y_{1:k}),
\end{equation}
where $\sigma_0=p(x_0)$. Now, substituting (\ref{simp_pos_den}) into (\ref{info}) yields
\begin{equation}\label{A2}
	\begin{split}
		\sigma_k
		&=\gamma_k p(y_k|x_k,x_{k-1})\\&\quad\times\int p(x_k|x_{k-1})\exp\big(\mu_{1,k-1}e_{k-1|k-1}^\top e_{k-1|k-1} \big)\\&\quad\times \exp\Big(\sum_{i=0}^{k-2}\mu_{1,i}e_{i|i}^\top e_{i|i}\Big)p(x_{k-1}|y_{1:k-1})dx_{k-1},\\
		&=\gamma_k p(y_k|x_k,x_{k-1}) \int p(x_k|x_{k-1})\\&\quad\times\exp\big(\mu_{1,k-1}e_{k-1|k-1}^\top e_{k-1|k-1} \big) \sigma_{k-1}dx_{k-1}.
	\end{split}
\end{equation}
We define the predicted information state density, $p(x_k|{\mathcal{I}_{k-1},e_{k-1|k-1})}$, as
\begin{equation}\label{Infor_den1}
	\begin{split}
		p(x_k|&\mathcal{I}_{k-1},e_{k-1|k-1})=\int p(x_k|x_{k-1})\\&\times\exp\big(\mu_{1,k-1}e_{k-1|k-1}^\top e_{k-1|k-1} \big)\sigma_{k-1}dx_{k-1}.
	\end{split}
\end{equation} 
From (\ref{A2}), we can write the posterior density for the information state as
\begin{equation}\label{Infor_den}
	\begin{split}
		\sigma_k\triangleq p(x_k|\mathcal{I}_k)&=\gamma_kp(y_k|x_k,x_{k-1})p(x_k|\mathcal{I}_{k-1},e_{k-1|k-1}).
	\end{split}
\end{equation}

Further, the cost function defined in (\ref{cost}) can be written as
\begin{equation*}
	\begin{split}
		J_k(\hat{x}_{k|k}|y_{1:k})= \int &\exp\left(\sum_{i=1}^{k-1}\mu_{1,i}e_{i|i}^\top e_{i|i} + \mu_{2,k} e_{k|k}^\top e_{k|k}\right)\\&\times p(x_k|y_{1:k})dx_k,
	\end{split}
\end{equation*}
and by (\ref{info}), the above equation reduces to 
\begin{equation}\label{final_cost}
	J_k(\hat{x}_{k|k}|y_{1:k})= \int \exp\left(\mu_{2,k} e_{k|k} e_{k|k}^\top\right)\sigma_kdx_k.
\end{equation}
{When the Eq.~(\ref{min_fun}) is solved recursively with the help of (\ref{Infor_den1})-(\ref{final_cost}), we receive risk sensitive estimates for the randomly delayed measurements.  
Note that the Eq.~(\ref{final_cost}) is an exponential quadratic cost function that explicitly considers only the present error. 
If the underlying system is linear and the noises are Gaussian, a closed form solution can be obtained. However, for a nonlinear system, the posterior density often becomes numerically intractable and usually, an approximate solution is approached. }

\textbf{Remark~2.}
Although $p(x_k|\mathcal{I}_k)$ is an unnormalized density, it doesn't change the value of estimate $\hat{x}_{k|k}$ for which (\ref{final_cost}) is minimum \cite{boel2002robustness}.

\textbf{Remark~3.} It can be observed that for a risk neutral case, $\mu_{1,k-1}=0$, and $\mu_{2,k}>0$, $p(x_k|\mathcal{I}_k)$ equals the $p(x_k|y_{1:k})$, and the cost function $J_k(\hat{x}_{k|k}|y_{1:k})$ reduces to a standard exponential quadratic function.
\section{ Risk sensitive filtering for a linear Gaussian system with randomly delayed measurements}\label{filter development}
{In this section, we use the general framework presented in previous section for a linear Gaussian system with randomly delayed measurements and derive a closed form solution of the estimate.} A discrete time linear system with model uncertainty is given by:
\begin{equation}\label{sys}
	x_k=(A_{k-1}+\Delta A_{k-1})x_{k-1}+w_{k-1},
\end{equation}
\begin{equation}\label{nd_lin_meas}
	z_k=C_kx_k+v_k,
\end{equation}
where $w_{k-1}$ and $v_k$ are zero mean, white Gaussian and mutually independent noise sequences with covariances $Q_{k-1}$ and $R_k$, respectively. The delayed measurement, $y_k$, is the same as defined in (\ref{meas}).  The initial state $x_0$ is also assumed to follow a Gaussian distribution and mutually independent of $w_{k-1}$, $v_k$ and $\beta_k$.  {$A_{k-1}$ is the matrix for transitioning  $x_{k-1}$ to $x_k$ and is considered to be invertible.}  {$\Delta A_{k-1}$ is an arbitrary, deterministic and unknown modeling uncertainty.}  We further assume that if $A_k^{-1}$ exists, then, $(A_k+\Delta A_k)^{-1}$ also exists for all permissible values of $\Delta A_k$. Moreover, the matrices $A_k$, $A_k+\Delta A_k$ and $C_k$ are assumed to be bounded for all $k \geq 0$, i.e. they hold the following inequalities:
\begin{equation}\label{ineq}
	\begin{split}
	   {0\leq} &{(A_k+\Delta A_k)^\top (A_k+\Delta A_k)\leq\kappa_1\textbf{I}},\\
	   &\text{and} \ 0\leq C_k^\top C_k\leq\kappa_2\textbf{I},
	\end{split}
\end{equation}
where $\kappa_1$ and $\kappa_2$ are real positive constants.

\textbf{Remark~4.}  {The $\Delta A_{k-1}$ represents the deviation of the process model from its nominal one. Note that, since $\Delta A_{k-1}$ is  unknown to the estimator, the estimator works only with the nominal process dynamics (i.e. $A_{k-1}$) for transitioning the states.}
%
\subsection{Risk sensitive estimate with delayed measurements}
In order to obtain the state estimate, $\hat{x}_{k|k}$,  for a linear and Gaussian system,  Eqs. (\ref{Infor_den1}) and (\ref{Infor_den}) are realized and the cost function $J_k(\hat{x}_{k|k}|y_{1:k})$ is minimized. We assume that $\sigma_k$ is an unnormalized Gaussian distribution, provided $\mu_{1,k-1}$ is a sufficiently small non-negative number \cite{boel2002robustness}. This assumption can be justified as $\sigma_0=p(x_0)$ follows the Gaussian distribution and at the end of this subsection, it is established that $\sigma_k$ is Gaussian if $\sigma_{k-1}$ is taken as a Gaussian density. The prior risk sensitive estimate for the linear system described by (\ref{sys}) and (\ref{meas}) is derived below.
\begin{thm} \label{lemma 1}
	The predicted mean and the error covariance for the system described by Eqs.~(\ref{sys}) and (\ref{meas}) are given by
	\begin{equation}\label{time_update}
		\begin{split}
			\hat{x}_{k|k-1}&=A_{k-1}\hat{x}_{k-1|k-1}, \\
			\Sigma_{k|k-1}&=A_{k-1}(\Sigma_{k-1|k-1}^{-1}-2\mu_{1,k-1}\mathbf{I})^{-1}A_{k-1}^\top+Q_{k-1},
		\end{split}
	\end{equation} 
{where $\Sigma_{k-1|k-1}$ is the posterior error covariance at the time step $k-1$.}
\end{thm}
\begin{pf*}{Proof.}
{To compute the predicted estimate, we use the framework of the predicted information state density given in Eq. (\ref{Infor_den1}). Considering $\sigma_{k-1}$ is a Gaussian density, we have
\begin{equation}\label{sigma}
	\begin{split}
	\sigma_{k-1}&=\gamma_{k-1|k-1}(2\pi)^{-n_x/2}|\Sigma_{k-1|k-1}|^{-1/2} \exp\Big(-\frac{1}{2}\times\\&(x_{k-1}-\hat{x}_{k-1|k-1})^\top\Sigma_{k-1|k-1}^{-1}(x_{k-1}-\hat{x}_{k-1|k-1})\Big).	
	\end{split}
\end{equation}
Substituting (\ref{sigma}) in (12) yields
\begin{equation*}
	\begin{split}
		&p(x_k|\mathcal{I}_{k-1},e_{k-1|k-1})\\&=\gamma_{k-1|k-1}(2\pi)^{-n_x/2}|\Sigma_{k-1|k-1}|^{
			-1/2}\int
		p(x_k|x_{k-1})\\&\times\exp\Big[-\dfrac{1}{2}\Big((x_{k-1}-\hat{x}_{k-1|k-1})^\top\Sigma_{k-1|k-1}
		^{-1}\\&\times
		(x_{k-1}-\hat{x}_{k-1|k-1}) + (x_{k-1}-
		\hat{x}_{k-1|k-1})^\top(-2\mu_{1,k-1}\textbf{I})\\&\times(x_{k-1}-\hat{x}_{k-1|k-1})\Big)\Big]dx_{k-1}.
	\end{split}
\end{equation*}
Using the distributive property of matrices on the terms inside the $\mbox{exp}[\cdot]$,  we can rewrite the above equation as
\begin{equation}\label{interm}
	\begin{split}
		&p(x_k|\mathcal{I}_{k-1},e_{k-1|k-1})\\&=\gamma_{k-1|k-1}(2\pi)^{-n_x/2}|\Sigma_{k-1|k-1}|^{
			-1/2}\int
		p(x_k|x_{k-1})\\&\times\exp[-\dfrac{1}{2}\big((x_{k-1}-\hat{x}_{k-1|k-1})^\top(\Sigma_{k-1|k-1}^{-1}-2\mu_{1,k-1}\textbf{I})\\&\times
		(x_{k-1}-\hat{x}_{k-1|k-1})\big)]dx_{k-1},
	\end{split}
\end{equation}
	where $\mu_{1,k-1}$ is a non-negative real number with $|\Sigma_{k-1|k-1}^{-1}-2\mu_{1,k-1}\textbf{I}|> 0$ or $2\mu_{1,k-1}\Sigma_{k-1|k-1}<\textbf{I}$ for every $k$. Clearly, the exponential part of (\ref{interm}) represents a Gaussian distribution and any factor required to make it a normalized distribution can be adjusted into the constant outside the integral. Eq.~(\ref{interm}) can be rewritten as
\begin{equation}\label{integrand}
	\begin{split}
	p&(x_k|I_{k-1},e_{k-1|k-1})=\gamma_{k|k-1} \int p(x_k|x_{k-1})\\&\times \mathcal{N}(x_{k-1}; \hat{x}_{k-1|k-1}, (\Sigma_{k-1|k-1}^{-1}-2\mu_{1,k-1}\textbf{I})^{-1})dx_{k-1}.
	\end{split}
\end{equation}
Using the assumed process dynamics given in (\ref{sys}), we have $p(x_k|x_{k-1})=\mathcal{N}(x_k; A_{k-1}x_{k-1}, Q_{k-1})$ \cite{ho1964bayesian} . Now, substituting this in (\ref{integrand}) and applying the Gaussian product theorem (see Theorem~2.1 of \cite{challa2011fundamentals}), we can write
\begin{equation}\label{gpt}
	\begin{split}
		p(x_k|I_{k-1},&e_{k-1|k-1})=\gamma_{k|k-1} \int \mathcal{N}(x_k; \mathbf{M}, \mathbf{S})\\& \times \mathcal{N}(x_{k-1}; \mathbf{M}_1, \mathbf{S}_1)dx_{k-1},
	\end{split}
\end{equation}
where \begin{eqnarray*}
	\begin{split}
		\mathbf{M}&=A_{k-1}\hat{x}_{k-1|k-1}, \\
		\mathbf{S}&=A_{k-1} (\Sigma_{k-1|k-1}^{-1}-2\mu_{1,k-1}\textbf{I})^{-1} A_{k-1}^\top+Q_{k-1}, \\
		G&=(\Sigma_{k-1|k-1}^{-1}-2\mu_{1,k-1}\textbf{I})^{-1} A_{k-1}^\top \mathbf{S}^{-1},\\
		\mathbf{M}_1&= \hat{x}_{k-1|k-1}+G(x_k-A_{k-1}\hat{x}_{k-1|k-1}),\\
		\mathbf{S}_1&=(\Sigma_{k-1|k-1}^{-1}-2\mu_{1,k-1}\textbf{I})^{-1}\\&\qquad - GA_{k-1}(\Sigma_{k-1|k-1}^{-1}-2\mu_{1,k-1}\textbf{I})^{-1}.
	\end{split}
\end{eqnarray*}
 The Gaussian distribution $\mathcal{N}(x_k; \cdot)$ is a function of $x_k$ and can be kept outside the integral.  Using the property of the normalized distribution,  $\int\mathcal{N}(x_{k-1}; \cdot)\mathrm{d}x_{k-1}=1$, 
 $p(x_k|\mathcal{I}_{k-1},e_{k-1|k-1})$ in (\ref{gpt}) can be obtained up to a normalizing constant as
\begin{equation}\label{predict}
	\begin{split}
	p&(x_k|\mathcal{I}_{k-1},e_{k-1|k-1})\sim \mathcal{N}\big(x_k;\ A_{k-1}\hat{x}_{k-1|k-1}, \\&A_{k-1} (\Sigma_{k-1|k-1}^{-1}-2\mu_{1,k-1}\textbf{I})^{-1} A_{k-1}^\top+Q_{k-1}\big).
	\end{split}
\end{equation}
The mean and covariance of the above distribution establish the expressions given in (\ref{time_update}).} \qed
\end{pf*}
To compute the posterior information state, we need to derive the conditional expectation, $\mathbb{E}[y_k|\mathcal{I}_{k-1},e_{k-1|k-1}]$, the covariance, $\Sigma^{yy}_k$, and the cross covariance, $\Sigma_k^{xy}$. To carry out these derivations, some useful relations are as follows:
\begin{eqnarray*}\label{rel}
	\begin{array}{ll}
	(i)&\mathbb{E}[w_{k-1}w_{k-1}^\top|\mathcal{I}_k]=\mathbb{E}[w_{k-1}^\top w_{k-1}]=Q_{k-1},\\\quad& \mathbb{E}[v_{k} v_{k}^\top|\mathcal{I}_k]=\mathbb{E}[v_{k} v_{k}^\top]=R_{k}\ \mbox{and} \ \mathbb{E}[w_i v_j^\top|\mathcal{I}_k]=0. \ \\
	(ii)&\mathbb{E}[(x_k-\hat{x}_{k|k-1})]=0.\\
	(iii)&\mathbb{E}_\beta[\beta_k^2]=\mbox{var}(\beta_k)+(\mathbb{E}_\beta[\beta_k])^2=\alpha_k, \mathbb{E}_\beta[(1-\beta_k)^2]  \\\quad& =1-\alpha_k \ \mbox{and} \ \mathbb{E}_\beta[\beta_k(1-\beta_k)]=0.\\
	(iv)&\mathbb{E}_\beta[(\beta_k-\alpha_k)]=0,\ \mathbb{E}_\beta[(\beta_k-\alpha_k)^2]=\mathbb{E}_\beta[((1-\beta_k) \\\quad&-(1-\alpha_k))^2]=\alpha_k(1-\alpha_k).
\end{array}
\end{eqnarray*}
\begin{eqnarray*}
	\begin{array}{ll}
	(v)& \mathbb{E}[e_{k|k-1}w_{k-1}^\top|\mathcal{I}_k]=\mathbb{E}[(A_{k-1}x_{k-1}+w_{k-1}\\\quad&-\hat{x}_{k|k-1}) w_{k-1}^\top|\mathcal{I}_k]=Q_{k-1}.\end{array}
\end{eqnarray*}
\begin{lem}\label{lemma_0}
	The conditional expectation of the measurement, $y_k$, is given as
	\begin{equation}\label{con_exp}
		\begin{split}
		\mathbb{E}[y_k|\mathcal{I}_{k-1},e_{k-1|k-1}]&=(1-\alpha_k)C_k\hat{x}_{k|k-1}\\&\quad+\alpha_kC_{k-1}\hat{x}_{k-1|k-1},
		\end{split}
	\end{equation} 
   and the conditional covariance of $y_k$, $\Sigma_k^{yy}$, is expressed as
	\begin{equation}\label{covY}
		\begin{split}
			\Sigma_k^{yy}&=(1-\alpha_k)C_k\Sigma_{k|k-1}C_k^\top+\alpha_k\theta_{k-1}\Sigma_{k|k-1}\theta_{k-1}^\top\\&+\alpha_kR_{k-1}+(1-\alpha_k)R_k-\alpha_k\theta_{k-1}Q_{k-1}\theta_{k-1}^\top+\alpha_k\\&\times(1-\alpha_k)(\theta_{k-1}-C_k) \hat{x}_{k|k-1}\hat{x}_{k|k-1}^\top(\theta_{k-1}-C_k)^\top,
		\end{split}
	\end{equation}
	where $\theta_{k-1}=C_{k-1}A_{k-1}^{-1}$.
\end{lem}
\begin{pf*}{Proof.}
	The measurement, $y_k$, is independent of the past errors $e_{1|1}, \cdots, e_{k-1|k-1}$. By using (\ref{meas}), we can write
	\begin{eqnarray*}
		\begin{split}
			\mathbb{E}[y_k|&\mathcal{I}_{k-1},e_{k-1|k-1}]\\&=\mathbb{E}[((1-\beta_k)(C_kx_k+v_k)\\&\qquad+\beta_k(C_{k-1}x_{k-1}+v_{k-1}))|y_{1:k-1}]\\
			&=\mathbb{E}_\beta[(1-\beta_k)]\mathbb{E}_x[C_kx_k+v_k] \\&\qquad+\mathbb{E}_\beta[\beta_k]\mathbb{E}_x[C_{k-1}x_{k-1}+v_{k-1}]\\
			&=(1-\alpha_k)C_k\hat{x}_{k|k-1}+\alpha_kC_{k-1}\hat{x}_{k-1|k-1}.
		\end{split}
	\end{eqnarray*}
Next, the covariance, $\Sigma_k^{yy}$, can be calculated as
\begin{equation}\label{sigma_y}
	\begin{split}
		\Sigma_k^{yy}&=\mathbb{E}\Big[\Big(y_k-\mathbb{E}[y_k|\mathcal{I}_{k-1},e_{k-1|k-1}]\Big)\\&\quad\times\Big(y_k-\mathbb{E}[y_k|\mathcal{I}_{k-1},e_{k-1|k-1}]\Big)^\top\Big]\\
		&=\mathbb{E}\Big[\Big((1-\beta_k)(C_kx_k+v_k)+\beta_k(C_{k-1}x_{k-1}+v_{k-1})\\&\quad-(1-\alpha_k)C_k\hat{x}_{k|k-1}-\alpha_kC_{k-1}\hat{x}_{k-1|k-1}\Big)\Big(\ast\Big)^\top\Big],
	\end{split}
\end{equation}
	where $(\ast)$ denotes the same terms given in the parenthesis left to it. Now, using the backward evolution of the state, $x_{k-1}=A_{k-1}^{-1}(x_k-w_{k-1})$, and rearranging the terms of (\ref{sigma_y}), we have
\begin{equation*}
	\begin{split}
		&\Sigma_k^{yy}\\&=\mathbb{E}\Big[\Big((1-\beta_k)C_k(x_k-\hat{x}_{k|k-1})+(1-\beta_k)C_k\hat{x}_{k|k-1}\\& +\beta_kC_{k-1}A_{k-1}^{-1}(x_k-\hat{x}_{k|k-1})+\beta_kC_{k-1}A_{k-1}^{-1}\hat{x}_{k|k-1}\\&-\beta_kC_{k-1}A_{k-1}^{-1}w_{k-1}+\beta_kv_{k-1}+(1-\beta_k)v_k-(1-\alpha_k)\\&\quad\times C_k\hat{x}_{k|k-1}-\alpha_kC_{k-1}A_{k-1}^{-1}\hat{x}_{k|k-1}\Big)\Big(\ast\Big)^\top\Big],
	\end{split}
\end{equation*}
\begin{equation}\label{sigma_y1}
	\begin{split}
	&\text{or,} \quad	\Sigma_k^{yy}\\&=\mathbb{E}\Big[\Big((1-\beta_k)C_ke_{k|k-1}+\beta_kC_{k-1}A_{k-1}^{-1}e_{k|k-1}+\beta_kv_{k-1}\\&\qquad+(1-\beta_k)v_k-\beta_kC_{k-1}A_{k-1}^{-1}w_{k-1}+(\beta_k-\alpha_k)\\&\qquad\times(C_{k-1}A_{k-1}^{-1}-C_k)\hat{x}_{k|k-1}\Big)\Big(\ast\Big)^\top\Big].
	\end{split}
\end{equation}
Now, multiplying the terms in the two parenthesis and applying the expectation operator with help of the relationships (i)-(v) mentioned before Lemma 2, Eq.~(\ref{sigma_y1}) reduces to (\ref{covY}). \qed
\end{pf*}
\begin{lem}\label{lemma_1}
	The cross covariance, $\Sigma_k^{xy}$, can be expressed as
	\begin{equation}\label{sigma_xy}
		\Sigma_k^{xy}=\Sigma_{k|k-1}[(1-\alpha_k)C_k+\alpha_k\theta_{k-1}]^\top-\alpha_kQ_{k-1}\theta_{k-1}^\top.
	\end{equation}
\end{lem}
\begin{pf*}{Proof.}
	By definition, we can write
	\begin{equation*}
		\begin{split}
			\Sigma_k^{xy}=\mathbb{E}\Big[\Big(x_k-\hat{x}_{k|k-1}\Big)\Big(y_k-\mathbb{E}[y_k|\mathcal{I}_{k-1},e_{k-1|k-1}]\Big)^\top\Big].
		\end{split}
	\end{equation*}
	Carrying out the operations similar to Lemma \ref{lemma_0} on the second term of the above expectation, we have
	\begin{equation}\label{sigma_xy_der}
		\begin{split}
			\Sigma&_k^{xy}=\mathbb{E}\Big[\Big(e_{k|k-1}\Big)\Big((1-\beta_k)C_ke_{k|k-1}+\beta_kC_{k-1}A_{k-1}^{-1}\\&\times e_{k|k-1}+\beta_kv_{k-1}+(1-\beta_k)v_k-\beta_kC_{k-1}A_{k-1}^{-1}w_{k-1}\\& \quad +(\beta_k-\alpha_k)(C_{k-1}A_{k-1}^{-1}-C_k)\hat{x}_{k|k-1}\Big)^\top\Big].
		\end{split}
	\end{equation}
	Again, rearranging the terms and using the relationship, $\mathbb{E}[e_{k|k-1}w_{k-1}^\top|\mathcal{I}_{k}]=Q_{k-1}$,  Eq. (\ref{sigma_xy_der}) can be further reduced as
	\begin{equation*}
		\begin{split}
			\Sigma_k^{xy}&=(1-\alpha_k)\Sigma_{k|k-1}C_k^\top+\alpha_k\Sigma_{k|k-1}(C_{k-1}A_{k-1}^{-1})^\top\\&\quad-\alpha_kQ_{k-1}\theta_{k-1}^\top\\&=\Sigma_{k|k-1}[(1-\alpha_k)C_k+\alpha_k\theta_{k-1}]^\top-\alpha_kQ_{k-1}\theta_{k-1}^\top.
		\end{split}
	\end{equation*}\qed
\end{pf*}
\begin{thm}\label{lemma 4}
	The posterior estimate and error covariance for the system described in (\ref{sys}) and (\ref{meas}) can be obtained as
	\begin{equation}\label{updated_est}
		\begin{split}
			\hat{x}_{k|k}&=\hat{x}_{k|k-1}+\Sigma_k^{xy}(\Sigma_k^{yy})^{-1}\big(y_{k}-(1-\alpha_k)C_k\hat{x}_{k|k-1}\\&-\alpha_kC_{k-1}\hat{x}_{k-1|k-1}\big),\\
			\Sigma_{k|k}&=\Sigma_{k|k-1}-\Sigma_k^{xy}(\Sigma_k^{yy})^{-1}\Sigma_k^{xy^\top}.
		\end{split}
	\end{equation} 
\end{thm}
\begin{pf*}{Proof.}
	The posterior information state, $p(x_k|\mathcal{I}_k)$, can be calculated in terms of the predicted information state,  $p(x_k|\mathcal{I}_{k-1},e_{k-1|k-1})$, and the current measurement, $y_k$. From (\ref{Infor_den}), we can write
	\begin{equation}\label{cond}
		\begin{split}
			p(x_k|\mathcal{I}_k)&=\gamma_kp(x_k,y_k|\mathcal{I}_{k-1},e_{k-1|k-1},x_{k-1})\\
			&=\gamma_kp(x_k,y_k|\mathcal{I}_{k-1},e_{k-1|k-1}),
		\end{split}
	\end{equation}
	where $x_{k-1}$ is the redundant information if $e_{k-1|k-1}$ is known.
	Assuming that $p(y_k|\mathcal{I}_{k-1},e_{k-1|k-1})$ is Gaussian and given as $p(y_k|\mathcal{I}_{k-1},e_{k-1|k-1}) \sim \mathcal{N}(y_k; \mathbb{E}[y_k|\mathcal{I}_{k-1}\\,e_{k-1|k-1}], \Sigma_k^{yy})$, where $\mathbb{E}[y_k|\mathcal{I}_{k-1},e_{k-1|k-1}]$ is derived in Lemma~\ref{lemma_0}. Also, by (\ref{time_update}) and (\ref{predict}), we can write $p(x_k\mathcal{I}_{k-1},e_{k-1|k-1}) \sim \mathcal{N}(x_k; \hat{x}_{k|k-1}, \Sigma_{k|k-1})$. Hence, writing for the joint Gaussian density, $p(x_k,y_k|\mathcal{I}_{k-1}\\,e_{k-1|k-1})$, we have
	\begin{equation}\label{joint_y}
		\begin{split}
		&p(x_k|\mathcal{I}_k)\\&=\gamma_{k}\mathcal{N}\left(\begin{bmatrix}y_k\\x_k\end{bmatrix};\begin{bmatrix}\mathbb{E}[y_k|\mathcal{I}_{k-1},e_{k-1|k-1}]\\\hat{x}_{k|k-1}\end{bmatrix}, \begin{bmatrix} \Sigma_k^{yy} & \Sigma_k^{yx}\\\Sigma_k^{xy} & \Sigma_k^{xx}\end{bmatrix}\right),
		\end{split}
	\end{equation}
	where the covariance, $\Sigma_k^{xx}=\Sigma_{k|k-1}$.
	Rearranging the terms of  (\ref{joint_y}) and performing the square operation for determination of the $p(x_k|\mathcal{I}_{k})$ as given in the Appendix~\ref{A}, we obtain
	\begin{equation}\label{cond2}
		\begin{split}
			&p(x_k|\mathcal{I}_k)=\gamma_{k|k}\exp\Big[-\frac{1}{2}\Big(x_k-\hat{x}_{k|k-1}-\Sigma_k^{xy}(\Sigma_k^{yy})^{-1}\\&\times(y_k-\mathbb{E}[y_k|\mathcal{I}_{k-1},e_{k-1|k-1}])\Big)^\top \Big(\Sigma_k^{xx}-\Sigma_k^{xy}(\Sigma_k^{yy})^{-1}\\&\times(\Sigma_k^{xy})^\top\Big)^{-1}\Big(x_k-\hat{x}_{k|k-1}-\Sigma_k^{xy}(\Sigma_k^{yy})^{-1}\\&\qquad\times(y_k-\mathbb{E}[y_k|\mathcal{I}_{k-1},e_{k-1|k-1}])\Big)\Big].
		\end{split}
	\end{equation}
 Eq.~(\ref{cond2}) is a Gaussian density, provided all the covariance matrices are invertible and $\mu_{1,k-1}$ is a sufficiently small positive number with $2\mu_{1,k-1}\Sigma_{k-1|k-1}<\textbf{I}$. Hence, the posterior estimate and error covariance are the mean and covariance of the density, $p(x_k|\mathcal{I}_k)$. Finally, Eqs.~(\ref{con_exp}) and (\ref{cond2}) establish (\ref{updated_est}).\qed
\end{pf*}
\textbf{Remark 5.} 
If  $\mu_{2,k}>0$ and $\mu_{1,k-1}\geq0$ with $2\mu_{1,k-1}\Sigma_{k-1|k-1} <\textbf{I}$,  $\hat{x}_{k|k}$ is the optimal estimate with respect to the cost function $J_k(\hat{x}_{k|k}|y_{1:k})$.

\textbf{Remark 6.}
It is straightforward to prove that if measurements are not delayed ({i.e.} $\alpha_k=0,\ \forall k$), the proposed filter converges to the RSKF given in \cite{boel2002robustness}, \cite{dey1997risk} and \cite{moore1995risk}.

\textbf{Remark~7.}
Note that if $\mu_{1,k-1}=0$ and $\alpha_k=0,\ \forall k$, the cost function (\ref{final_cost}) becomes a standard exponential-quadratic function with no past error, and the proposed filter coincides with the Kalman filter.

\textbf{Remark~8.}
If we rewrite the posterior estimate as $\hat{x}_{k|k}=L_k\hat{x}_{k|k-1}+K_ky_k$, where $L_k=\textbf{I}-K_k\big((1-\alpha_k)C_k+\alpha_kC_{k-1}A_{k-1}^{-1}\big)$ and $K_k=\Sigma_k^{xy}(\Sigma_k^{yy})^{-1}$, then, using Proposition 1 of \cite{ray1993state}, the proposed estimator can be shown unbiased, i.e. $\mathbb{E}[x_k-\hat{x}_{k|k}]=\mathbb{E}[e_{k|k}]=0$, provided $\mu_{2,k}>0$, and $\mu_{1,k-1}$ is a small non-negative real number with $2\mu_{1,k-1}\Sigma_{k-1|k-1}<\textbf{I},\ \forall k$.
\subsection{Selection of risk sensitive parameter}
 In previous works  \cite{dey1997risk}, \cite{boel2002robustness}, the risk sensitive parameters are assumed to be  time invariant, however, such restriction is unnecessary and we relaxed that assumption. The risk sensitive parameter, $\mu_{2,k}$, does not affect the optimization of the cost function (\ref{final_cost}) as long as it is a positive real number. Also, the expressions for the posterior estimate and covariance are independent of $\mu_{2,k}$. But, the selection of the risk sensitive parameter, $\mu_{1,k-1}$, is of main concern as it determines the contribution of past errors in calculating the current estimate. From Eq. (\ref{interm}), imposing the fact that the predicted covariance, $\Sigma_{k|k-1}$, must be positive definite, the parameter $\mu_{1,k-1}$ must satisfy $|\Sigma_{k-1|k-1}^{-1}-2\mu_{1,k-1}\textbf{I}|< 0$. Further, solving the inequality, we obtain an upper limit of $\mu_{1,k-1}$ and then select a value of it by keeping a safe tolerance from the upper limit to avoid the ill-conditioning of the resultant covariance. Note that calculating  $\mu_{1,k-1}$ at each time step increases the computational cost slightly.
\section{Stability of the delayed risk sensitive Kalman filter}
{At first, we establish the stability criteria for the RSKF in the mean square sense and then utilize it to comment on the stability of the proposed filter.}
	\subsection{Stability of the risk sensitive Kalman filter}
	The RSKF is an optimal filter \cite{boel2002robustness}, however, the optimality doesn't imply the stability \cite{jazwinski2007stochastic}. We assume that the state of the nominal system is bounded and define the observability and controllability matrices for the stochastic system (\ref{sys}), (\ref{nd_lin_meas}) as \cite{jazwinski2007stochastic}	
%
	 \begin{eqnarray}\label{onc}
	 	\begin{split}
	 		\mathcal{O}_{k,k-l}&=\sum_{i=k-l}^k (A_{i,k}+\Delta A_{i,k})^\top C_i^\top R_i^{-1}  C_i (A_{i,k}+\Delta A_{i,k})\\
	 		\mathcal{C}_{k,k-l}&=\sum_{i=k-l}^{k-1} (A_{k,i+1}+\Delta A_{k,i+1})^\top Q_i\\&\qquad\qquad\times (A_{k,i+1}+\Delta A_{k,i+1}); \ \forall k\geq l,
	 	\end{split}
	 \end{eqnarray}
	 where $l$ is a positive integer. $(A_{i,k}+\Delta A_{i,k})$ is the backward transition matrix for transitioning of the state from time step $k$ to $i$, and $(A_{k,i}+\Delta A_{k,i})$ is the forward transition matrix for transitioning the state from time step $i$ to $k$. The transition matrices for the actual system (\ref{sys}) are defined as below:
	 \begin{equation}\label{trans_mat}
	 	\begin{split}
	 		(A_{i,k}+\Delta A_{i,k})&=\prod_{j=1}^{k-i}(A_{k-j}+\Delta A_{k-j})^{-1},\\
	 		(A_{k,i}+\Delta A_{k,i})&=\prod_{j=i}^{k-1}(A_j+\Delta A_j); \quad 0\leq i <k,
	 	\end{split}
	 \end{equation}
 where $A_{k-1,k}=A_{k,k-1}^{-1}=A_{k-1}^{-1}, \ \Delta A_{i,k} =\prod_{j=1}^{k-i}(A_{k-j}+\Delta A_{k-j})^{-1} - \prod_{j=1}^{k-i}A_{k-i}^{-1}, \ \Delta A_{k,i}=\prod_{j=i}^{k-1}(A_j+\Delta A_j) - \prod_{j=i}^{k-1}A_j$, and $A_{k,k}=(A_{k,k}+\Delta A_{k,k})=\mathbf{I}$. Now, the system described by (\ref{sys}) and (\ref{nd_lin_meas}) is said to be uniformly completely observable and uniformly completely controllable if the observability matrix, $\mathcal{O}_{k,k-l}$, and controllability matrix, $\mathcal{C}_{k,k-l}$, are positive definite and bounded\cite{jazwinski2007stochastic}, i.e.
 \begin{eqnarray}\label{obs}
 	\begin{split}
 		&0<\kappa_3\textbf{I}\leq\mathcal{O}_{k,k-l}\leq\kappa_4\textbf{I},\\
 		& 0<\kappa_5\textbf{I}\leq\mathcal{C}_{k,k-l}\leq\kappa_6\textbf{I},
 	\end{split}
 \end{eqnarray}
 where $\kappa_3, \kappa_4, \kappa_5$, and $\kappa_6$ are real positive constants.
 
 \textbf{Remark~9.} {Since the nominal system is a special case of the actual system (when $\Delta A_k=0$), the observability matrix and the controllability matrix of the nominal system, represented by $O_{k,k-l}$ and $\cup_{k,k-l}$, respectively, are also positive definite and bounded if $\mathcal{O}_{k,k-l}$ and $\mathcal{C}_{k,k-l}$ are positive definite and bounded.}
 
{Now, we represent the posterior  error covariance of the RSKF with $\mathbf{P}_{k|k}$ and it is needless to mention that it will remain positive definite \cite{jazwinski2007stochastic}, i.e. $\mathbf{P}_{k|k}>0, \ k\geq 0$, provided $\mathbf{P}_0>0$ and $2\mu_{1,k-1}\mathbf{P}_{k-1|k-1}<\mathbf{I}$. 
The results of stability analysis are presented in following theorem.
 \begin{thm}\label{stabality}
 	If the system (\ref{sys}), (\ref{nd_lin_meas}) is uniformly completely observable and uniformly completely controllable, and if $\mathbf{P}_0>0$, and $2\mu_{1,k-1}\mathbf{P}_{k-1|k-1}<\mathbf{I}$, then $\mathbf{P}_{k|k}$ is uniformly bounded for all $k\geq l$, provided $-\mathbf{I}<O_{k,k-l}^{-1}\Delta O_{k,k-l}<\mathbf{I}$, where
 	$$\Delta O_{k,k-l}=\sum_{i=k-l}^k A_{i,k}^\top C_i^\top R_i^{-1} C_i\Delta A_{i,k}.$$
 \end{thm}
\begin{pf*}{Proof.}
	Please see Appendix~\ref{stability_proof}.\qed
	\end{pf*}}
\textbf{Remark~10.} { One of the possible cases where $\mathbf{P}_{k|k}$ tends to infinity (which means the RSKF becomes unstable) is when the magnitude of $\Delta O_{k,k-l}$ is greater or equal to that of $O_{k,k-l}$, i.e., the uncertainty in the transition matrix, $\Delta A_{j,k}\geq A_{j,k}$ for $\Delta A_{j,k}>0$ or, $\Delta A_{j,k} \leq -A_{j,k}$ for $\Delta A_{j,k}<0$.} 

\textbf{Remark~11.} {If there is no uncertainty in the model, i.e. if $\Delta O_{k,k-l}=0$, the inequality in (\ref{final_ineq}) reduces to $\mathbf{P}_{k|k}\leq O_{k,k-l}^{-1}+\cup_{k,k-l}$ as established in \cite{jazwinski2007stochastic} for the Kalman filter.
\subsection{Stability of the risk sensitive Kalman filter with randomly delayed measurements} 
 For a system with non-delayed measurements, i.e. $\alpha_k=0$, the proposed filter reduces to the RSKF and in Theorem~\ref{stabality} we showed that it is stable under certain conditions if the system is uniformly completely controllable and observable. It is obvious that the delay in measurements does not alter the controllability of the system, therefore, the delayed risk sensitive filter will be stable if the observability of the system is preserved in presence of random delay in measurements. }
	
{	Now, we augment the state of the system (\ref{sys}) with previous step state vector, i.e. $X_k= [x_k^\top \ x_{k-1}^\top]^\top$, and the augmented system becomes
	\begin{equation}\label{aug_sys}
		\begin{split}
			X_k&=(\phi_{k,k-1}+\Delta \phi_{k,k-1})X_{k-1}+W_{k-1},\\
			y_k&=\mathbf{C}_k +v'_k,
		\end{split}
	\end{equation} 
where\\ 
$\begin{array}{ll}
	\phi_{k,k-1}+\Delta \phi_{k,k-1}=\begin{bmatrix}A_{k-1}+\Delta A_{k-1}& 0\\ 0&A_{k-2}+\Delta A_{k-2} \end{bmatrix},\end{array}$ 
	$W_{k-1}=[w_{k-1}^\top w_{k-2}^\top]^\top, \ \mathbf{C}_k=[(1-\beta_k)C_k \ \beta_kC_{k-1}],$ and $v'_k=(1-\beta_k)v_k+\beta_kv_{k-1}$.  The covariance of the modified noise can be calculated as $R'_k=\mathbb{E}[v'_k{v'_k}^\top]=(1-\alpha_k)R_k+\alpha_k R_{k-1}.$ Clearly, the value of $R'_k$ lies between $R_{k-1}$ and $R_{k}$ for any $\alpha_k \in [0,1]$. Note that at a given time step $k$, there is $\alpha_{k-1}(1-\alpha_k)$ probability that $v'_k$ is not white and it becomes so when the filter use the same measurement data at two consecutive time steps. In such a scenario, the notion of observability defined in (\ref{onc}) is violated and a formal proof of the stability of the proposed filter is far to achieve. Therefore, we justify the stability in the form of a conjecture where we assume the noise, $v'_k$, is white.  
\begin{conj}
	If the system (\ref{meas}), (\ref{sys})  is uniformly completely observable with no delay in measurements ($\alpha_k=0$), then, the underlying system is also uniformly completely observable for $\alpha_k \in [0, 1-\epsilon]$, where $0<\epsilon\leq1$.
\end{conj}
\begin{pf*}{Justification.}%
	Assuming the modified noise, $v'_k$, is white, the observability matrix for the system (\ref{meas}), (\ref{sys}) is defined as
	\begin{equation}\label{obsG}
		\begin{split}
				\mathbf{O}_{k,k-l}=&\sum_{i=k-l}^k\mathbb{E}_\beta[(\phi_{i,k}+\Delta \phi_{i,k})^\top\mathbf{C}_i^\top {R'_i}^{-1} \mathbf{C}_i\\&\qquad\times(\phi_{i,k}+\Delta \phi_{i,k})].
				\end{split}
	\end{equation}
 Now, using (\ref{aug_sys}) and the relationship~(iii) given in Section~\ref{filter development}, we write
	$\mathbf{O}_{k,k-l}=\begin{bmatrix} \mathbf{O}_{11} & 0\\0 &\mathbf{O}_{22} \end{bmatrix},$
where 
\begin{equation*}\label{o11}
	\begin{split}
	\mathbf{O}_{11}&=\sum_{i=k-l}^k (1-\alpha_i)(A_{i,k}+\Delta A_{i,k})^\top{C}_i^\top {R'_i}^{-1} {C}_i\\&\qquad \times(A_{i,k}+\Delta A_{i,k}),\\
	\mathbf{O}_{22}&=\sum_{i=k-l}^k\alpha_i (A_{i-1,k}+\Delta A_{i-1,k})^\top{C}_{i-1}^\top {R'_{i-1}}^{-1}\\&\qquad\times{C}_{i-1}(A_{i-1,k}+\Delta A_{i-1,k}).\\
	\end{split}
\end{equation*}
Since we seek to find the observability of $x_k$ in the augmented state, $X_k=[x_k^\top \ x_{k-1}^\top]^\top$,  only $\mathbf{O}_{11}$ of $\mathbf{O}_{k,k-l}$ needs to be established as a positive definite and finite matrix \cite{chen1999linear}. Given that $\alpha_j \in [0, 1-\epsilon]$, the bounds of $\mathbf{O}_{11}$ can be expressed as 
\begin{equation}
	\epsilon \mathcal{O'}_{k,k-l} \leq \mathbf{O}_{11} \leq \mathcal{O'}_{k,k-l},
\end{equation}
 where $\mathcal{O'}_{k,k-l}$ is the observability matrix of (\ref{onc}) with the covariance, $R'_i$. Since $R'_i$ always lies in between $R_{i-1}$ and $R_i$, the observability matrix, $\mathcal{O'}_{k,k-l}$, follows the same bound as given in (\ref{obs}). Hence, $\mathbf{O}_{11}$ is a positive definite and bounded matrix for $0<\epsilon\leq 1$, and the system (\ref{meas}), (\ref{sys}) is uniformly completely observable. \qed
\end{pf*}
\textbf{Remark~12.} If $\epsilon=0$ (i.e. $\alpha_k=1$), which means all the measurements are one step delayed, the observability matrix, $\mathbf{O}_{11}$, becomes positive semidefinite, and hence, the system is not uniformly completely observable.}
\section{Estimation of latency probability}
In practice, the latency probability, $\alpha_k$, can be unknown to the user for a given system. In such a case, it must be identified before the estimation. In this section, we present a method to estimate the latency parameter assuming it is stationary i.e., $\alpha_k=\alpha,\ \forall k$, using the maximum likelihood (ML) criterion on the received measurements. It involves the maximization of the joint density $p_\alpha(y_{1:m})$ with respect to the latency parameter $\alpha$, which can be represented as \cite{zhang2016particle,tiwari2020particle}
\begin{equation*}
	\hat{\alpha}=\arg \underset{\alpha\in[0,1]}{\max} p_\alpha(y_1,\cdots,y_m), 
\end{equation*}
where $m$ is the total number of measurements used to estimate the latency parameter. Using the chain rule, the above joint density can be rewritten as
\begin{equation}\label{likelihood}
	p_\alpha(y_1,\cdots,y_m)=p(y_1)\prod_{k=2}^m p_\alpha(y_k|y_{1:k-1}),
\end{equation}
where the first received measurement, $y_1$, is considered to be non-delayed and independent of $\alpha$. For computational simplicity, we can take the logarithmic of (\ref{likelihood}) as below:
\begin{align}\label{log_likeli}
	\begin{split}
		L_\alpha(y_{1:m})&=\log p_\alpha(y_{1:m})\\
		&=\log p(y_1)+\sum_{k=2}^m\log p_\alpha(y_k|y_{1:k-1}).
	\end{split}
\end{align}
Considering the fact that the current measurement, $y_k$, is correlated with both the current state $x_k$ and the previous state $x_{k-1}$, and using the Bayes' Theorem for the likelihood $p_\alpha(y_k|y_{1:k-1})$ of (\ref{log_likeli}), we can write
\begin{equation}\label{meas_likelihood}
	\begin{split}
		p_\alpha(y_k|y_{1:k-1})=&\int \int p_\alpha(y_k|x_k,x_{k-1})p(x_k|x_{k-1})\\&\times p_\alpha(x_{k-1}|y_{1:k-1})dx_kdx_{k-1}.
	\end{split}
\end{equation}
Again, rewriting the received measurement as
\begin{equation}
	\begin{split}\label{mod_meas}
		y_k&=(1-\beta_k)h_k(x_k)+\beta_kh_{k-1}(x_{k-1})\\&\quad+(1-\beta_k)v_k+\beta_kv_{k-1}\\
		&=\psi_k(x_k,x_{k-1})+v'_k,
	\end{split}
\end{equation}
where $\psi_k(.)=(1-\beta_k)h_k(x_k)+\beta_kh_{k-1}(x_{k-1})$, and $v'_k=(1-\beta_k)v_k+\beta_kv_{k-1}$.
Considering that the modified measurement noise $v'_k$, conditioned on $v_k$ and $v_{k-1}$, is an independent sequence over time, and by using (\ref{mod_meas}), the state likelihood, $p_\alpha(y_k|x_k,x_{k-1})$, can be defined as
\begin{equation}\label{likeli_pdf}
	p_\alpha(y_k|x_k,x_{k-1})=p_{v'_k}(y_k-\psi_k(x_k,x_{k-1})),
\end{equation}
where $p_{v'_k}(\cdot)$ is the pdf for the modified measurement noise. Using the marginalization of the joint pdf of $v'_k$ and $\beta_k$, $p_{v'_k}(\cdot)$ can be constructed as
\begin{equation}\label{pdf_vk}
	\begin{split}
		p(v'_k)&=p(v'_k,\beta_k=0)+ p(v'_k,\beta_k=1)\\
		&=p(v'_k|\beta_k=0)p(\beta_k=0)+p(v'_k|\beta_k=1)p(\beta_k=1).
	\end{split}
\end{equation}
From the eqn. (\ref{mod_meas}), we can write $p(v'_k|\beta_k=0)=p(v_k)$ and $p(v'_k|\beta_k=1)=p(v_{k-1})$. Hence, the eqn. (\ref{pdf_vk}) reduces to
\begin{equation}\label{pdf_vk1}
	p(v'_k)=(1-\alpha_k)p_{v_k}+\alpha_k p_{v_{k-1}}.
\end{equation}
Substituting (\ref{pdf_vk1}) into (\ref{likeli_pdf}), we can rewrite the Eq. (\ref{likeli_pdf}) as
\begin{equation}\label{state_likelihood}
	\begin{split}
		p_\alpha(y_k|x_k,x_{k-1})&=(1-\alpha_k)p_{v_k}(y_k-h_k(x_k))+\alpha_kp_{v_{k-1}}\\&\times(y_k-h_{k-1}(x_{k-1})).
	\end{split}
\end{equation}
Now, the likelihood function of (\ref{meas_likelihood}) can be rewritten as
\begin{equation}\label{meas_likelihood1}
	\begin{split}
		p_\alpha(y_k|y_{1:k-1})&=\int \int p_\alpha(y_k|x_k,x_{k-1})\\& \times p_\alpha(x_k,x_{k-1}|y_{1:k-1})dx_kdx_{k-1}\\
		&=\mathbb{E}[p_\alpha(y_k|x_k,x_{k-1})].
	\end{split}
\end{equation}
Substituting (\ref{meas_likelihood1}) into the log likelihood expression in (\ref{log_likeli}) and using (\ref{state_likelihood}) yields
\begin{equation}\label{final_ll}
	\begin{split}
		L_\alpha(y_{1:m})=&\sum_{k=2}^m\log\big[\mathbb{E}[(1-\alpha_k)p_{v_k}(y_k-h_k(x_k))\\&\quad+\alpha_kp_{v_{k-1}}(y_k-h_{k-1}(x_{k-1}))]\big],
	\end{split}
\end{equation}
where the measurement $y_1$ is ignored as it is not the function of parameter $\alpha$. Since in (\ref{final_ll}), both the latency parameter and the states are unknown, the analytical maximization of the log likelihood is certainly complicated, and thus approximation is necessary. Eq. (\ref{final_ll}) can be maximized numerically over $\alpha \in [0,1]$ while we use sequential Monte Carlo (SMC) approximation for the computation of log likelihood.
\subsection{Computation of log likelihood}
Considering the SMC approximation to compute the expectation, we can write the log likelihood function for the linear system as
\begin{eqnarray}\label{LL_f}
	\begin{split}
		L_\alpha(y_{1:m})=\sum_{k=2}^m&\log\big[\dfrac{1}{N}\sum_{i=1}^N(1-\alpha_k)p_{v_k}(y_k-C_kx_k^i)\\&+\alpha_kp_{v_{k-1}}(y_k-C_{k-1}x_{k-1}^i)\big],
	\end{split}
\end{eqnarray}
where $x_{k-1}^i$ are sampled from the density, $p_\alpha(x_{k-1}|y_{1:k-1})\\=\mathcal{N}(x_{k-1};\hat{x}_{k-1|k-1},\Sigma_{k-1|k-1})$, $x_k^i$ are sampled from the density, $p(x_k|x_{k-1})=\mathcal{N}(x_{k};A_{k-1}\hat{x}_{k-1|k-1},A_{k-1}\times\\(\Sigma_{k-1|k-1}^{-1}-2\mu_{1,k-1}\textbf{I})^{-1}A_{k-1}^\top+Q_{k-1})$, and $N$ is the total number of samples.
Algorithm 1 outlines the steps that can be used to estimate the latency parameter.
\begin{algorithm}
	\caption{Estimation of latency parameter}
	\begin{enumerate}
		\item Select the values for step length ($l$), and number of measurements ($m$).
		\item Calculate the log likelihood, $L_\alpha$ for $\alpha=0$, and set $L_{max}=L_0$.
		\item \textit{for} $\alpha=0:l:1$
		\begin{itemize}
			\item \textit{for} $k=2:m$
			\begin{itemize}
				\item Calculate the posterior estimate, $\hat{x}_{k-1|k-1}$ and covariance, $\Sigma_{k-1|k-1}$ using eq. (\ref{time_update}) and (\ref{updated_est}).
				\item Sample $x_{k-1}^i \sim$$\mathcal{N}(x_{k-1};\hat{x}_{k-1|k-1},\Sigma_{k-1|k-1})$
				\item Resample $x_{k-1}^i$ according to its likelihood and obtain $\bar{x}_{k-1|k-1}^i$.
				\item Compute $x_k^i=A_{k-1}\bar{x}_{k-1|k-1}$.
				\item Calculate the state likelihood, $p_\alpha(y_k|x_k,x_{k-1})$ using the SMC approximation and (\ref{state_likelihood}).
				\item Calculate $L_\alpha=L_\alpha+\log (p_\alpha(y_k|x_k,x_{k-1}))$
			\end{itemize}
			\item \textit{end for}
			\item \textit{if} $L_\alpha>LL$
			\begin{itemize}
				\item $L_{max}=L_\alpha$, $\hat{\alpha}=\alpha$
			\end{itemize}
			\item \textit{end if}
		\end{itemize}
		\item \textit{end for}
	\end{enumerate}
\end{algorithm}
\section{Simulation Results}
In this section, two numerical problems are simulated to demonstrate the effectiveness of the proposed filter over the existing RSKF \cite{boel2002robustness} and the KF-RD \cite{ray1993state}. Considering the stationary statistics for the random delays in measurements, first, the latency probability is estimated by maximizing the likelihood described in Eq.~(\ref{LL_f}) over $\alpha \in [0,1]$. Subsequently, the estimated latency probability, $\alpha$, is used in implementing the proposed filter for given problems.
\subsection{Problem: 1}
Consider a two dimensional linear stochastic system \cite{xie1994robust}
\begin{equation}\label{prob1}
	\begin{split}
		x_k&=\begin{bmatrix}0& -0.5\\1& 1\end{bmatrix}x_{k-1}+\begin{bmatrix}
			-6\\1 \end{bmatrix}w_{k-1},\\
		z_k&=[-10 \ 1]x_k+v_k,
	\end{split}
\end{equation}
where $w_{k-1}$ is a white Gaussian sequence with zero mean and unity covariance, $v_k$ is a white Gaussian sequence with zero mean and covariance, $R=3.6$. $w_{k-1}$ and $v_k$ are uncorrelated sequences. The uncertainty in system modeling is represented by $\Delta A=\begin{bmatrix}0&0\\0 &\delta \end{bmatrix}.$ The truth is initialized with $x_0 \sim \mathcal{N}(0,\Sigma_0)$, where $\Sigma_0=\begin{bmatrix}1 &0\\0&5\end{bmatrix}$.

\begin{figure}
	\centering
	\includegraphics[width=3.2in,height=4cm]{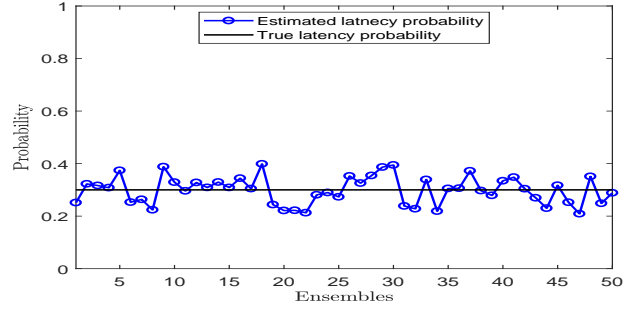}
	\caption{ Estimated latency probability.}
	\label{lat_1}
\end{figure}

\begin{figure}
	\centering
	\includegraphics[width=3.2in,height=4cm]{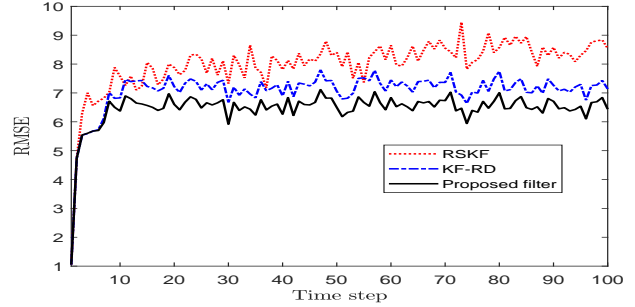}
	\caption{ {RMSE of state-1 with $\delta=0.35$ and $\alpha=0.60$.}}
	\label{rmse_x1}
\end{figure}

\begin{figure}
	\centering
	\includegraphics[width=3.2in,height=4cm]{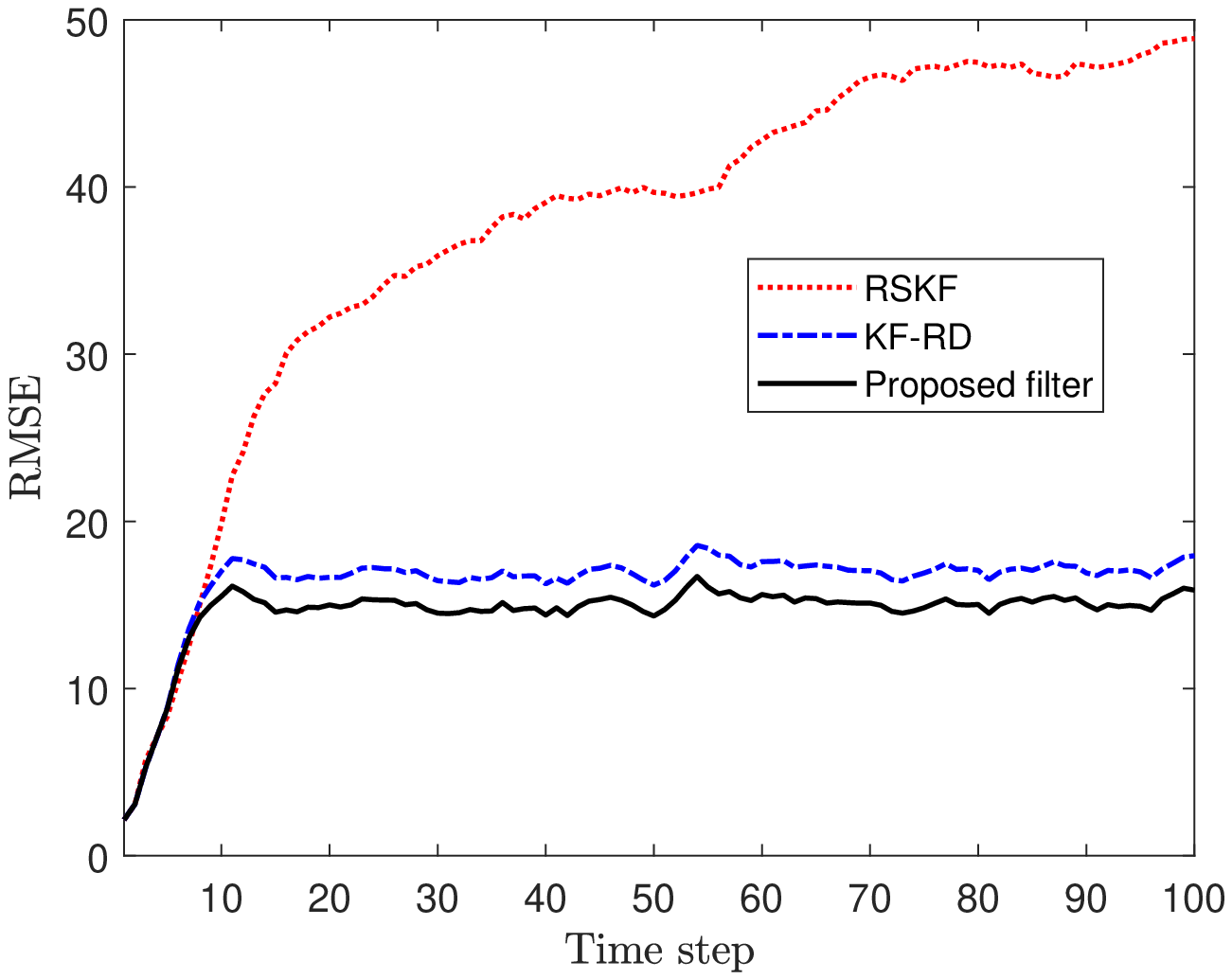}
	\caption{ {RMSE of state-2 with $\delta=0.35$ and $\alpha=0.60$.}}
	\label{rmse_x2}
\end{figure}

 Assuming that the measurement delay statistics is stationary, i.e. $\mathbb{E}[\beta_k]=\alpha, \ \forall k$, the estimation of latency parameter (with the help of Algorithm 1) is carried out for each ensemble and plotted in Fig.~\ref{lat_1}. From the figure it can be seen that at each ensemble the estimated value is near to its truth. The average of the estimated values over $50$ ensembles is 0.291 when the true latency parameter considered is $0.3$.
   The proposed filter is implemented along with the estimated value of the latency probability and its performance is compared with that of the RSKF and the KF-RD. The metrics used for evaluating the performance of different filters are the root mean square error (RMSE) and the time averaged mean square error (Avg-MSE), which are calculated over 500 Monte Carlo runs. The risk sensitive parameter, $\mu_{1,k-1}$, is selected such that its value is less than the real and positive roots of the  equation $|\Sigma_{k-1|k-1}^{-1}-2\mu_{1,k-1}\textbf{I}|= 0$.
Figs. \ref{rmse_x1} and \ref{rmse_x2} show the RMSE of states when the uncertainty is taken, $\delta=0.35$, and the true latency probability, $\alpha=0.5$. It can be seen that the proposed filter outperforms the other two filters in presence of the modeling uncertainty and the random delay in measurements. 
\begin{table}
\begin{center}
	\caption{The Avg-MSE of different filters for various $\alpha$ and $\delta$.}\label{Tab:1}
	\begin{tabular}{|p{0.4cm}|p{1.15cm}||p{0.7cm}|p{0.8cm}|p{0.73cm}|p{0.78cm}|p{0.78cm}|}
		\hline
		\centering \multirow{2}{*}{$\delta$}&\multirow{2}{*}{}& \multicolumn{5}{c|}{$\alpha$}\\[1ex]
		\cline{3-7}\centering
		\centering&&$0$&$0.20$&$0.40$&0.60&$0.80$\\[1.2ex]
		\hline\hline
		\centering	\multirow{3}{*}{$0$}&RSKF&0.072 &17.872 &34.025 &49.263 &64.474  \\[1ex]
		\cline{2-7}
		&KF-RD&0.068 &13.217 &23.102 &31.241 &36.889 \\[1ex]
		\cline{2-7}
		&Proposed filter &0.072 &13.294 &23.332 &31.636 &37.616\\[2ex]
		\hline
		\multirow{3}{*}{$0.15$}&RSKF&0.724 &18.282 &34.322 &50.109 &64.203  \\[1ex]
		\cline{2-7}
		&KF-RD&1.008 &13.393 &24.201 &33.102 &40.001   \\[1ex]
		\cline{2-7}
		&Proposed filter &0.724 &13.171 &23.710 &32.510 &39.141  \\
		\hline
		\multirow{3}{*}{$0.25$}&RSKF&3.130 &20.901 &37.007 &53.416 &67.712  \\[1ex]
		\cline{2-7}
		&KF-RD&4.880 &15.152 &26.910 &37.713 &48.360   \\[1ex]
		\cline{2-7}
		&Proposed filter &3.130 &14.105 &25.001 &35.130 &43.925 \\[2ex]
		\hline
		\multirow{3}{*}{$0.35$}&RSKF&13.802 &30.601 &48.404 &64.205 &80.101  \\[1ex]
		\cline{2-7}
		&KF-RD&23.401 &20.681 &35.593 &52.791 &71.702   \\[1ex]
		\cline{2-7}
		&Proposed filter &13.802 &16.901 &29.595 &51.893 &55.796  \\[2ex]
		\hline
	\end{tabular}
\end{center}
\end{table}

{ A parametric study is carried out by varying the uncertainty parameter, $\delta$, and the probability, $\alpha$. Table \ref{Tab:1} displays the Avg-MSE of state-1 from the different filters for various set of the parameters. Similar results are obtained for state-2 and not shown here. Some observations that can be made from this parametric study are as follows:}
\begin{enumerate}
	\item[i.] {Without any uncertainty in process dynamics ($\delta=0$), the proposed filter with a nonzero risk parameter performs better than the RSKF and is comparable to the KF-RD. This follows the fact that the risk sensitive parameter, $\mu_{1,k-1}$, which scales the accumulated past errors, is not set to zero in the RSKF, despite there being no need to minimize the exponential of past errors, whereas, the KF-RD works in a risk neutral way.} 
	\item[ii.] With the increase in $\alpha$, the Avg-MSE increases for all the filters.
	\item[iii.] It is also to be observed that in presence of the random delay in measurements, the KF-RD, which is designed to handle the random delays, is more effective than the RSKF.
		\item[iv.] {In the presence of uncertainty in the process model, the proposed filter always performs better than the other two filters irrespective of the value of delay probability, $\alpha$. Also, the improvement in the performance of the proposed estimator over the other filters becomes more prominent when the actual process dynamics deviate more from the nominal one, and the random delay in measurements is more likely.}
\end{enumerate}
\subsection{Problem: 2}
{Consider a constant turn rate model for an aircraft that executes a maneuvering turn in a two dimensional plane with a fixed, but uncertain turn rate $\Omega$.  The four dimensional state vector for the kinematics of aircraft is considered as $x_k=[\eta_k\ \nu_k\ \dot{\eta}_k\ \dot{\nu}_k]^\top$, where $\eta_k$ and $\nu_k$ represent positions, and $\dot{\eta}_k$ and $\dot{\nu}_k$ represent velocities along the $X$ and $Y$ coordinates, respectively. The discrete time system model is given as}
\begin{equation*}
	\begin{split}
		x_k&=A_{k-1}x_{k-1}+Bw_{k-1}\\
		z_k&=C_kx_k+v_k,
	\end{split}
\end{equation*}
where \begin{equation*}
	\begin{split}
		A_{k-1}=\begin{bmatrix}
			1 &0 & \dfrac{\sin \Omega T}{\Omega} & \dfrac{\cos \Omega T-1}{\Omega}\\
			0 &1& \dfrac{1-\cos \Omega T}{\Omega}& \dfrac{\sin \Omega T}{\Omega}\\
			0 & 0&\cos \Omega T & -\sin \Omega T\\
			0 &0& \sin \Omega T & \cos \Omega T\end{bmatrix},
	\end{split}
\end{equation*} 
$T$ is sampling period, and the noise gain, $B=aT^2[\frac{1}{2}\mathbf{I}_{2\times2} \ \mathbf{I}_{2\times2}]^\top$.
The measurement is given as the noise corrupted  $X$ and $Y$ coordinate of the target, therefore, $C_k=[\mathbf{I}_{2\times2} \ 0_{2\times2}]$.
$w_{k-1}$ and $v_k$ are uncorrelated zero mean white Gaussian sequences with covariances $ Q=\text{diag}([0.3^2\ 0.3^2\ 0.05^2])$ and $R=\text{diag}([12\ 12])$, respectively. Initial values are taken as $x_0=[200\ 200\ 15\ 15]^\top$, and {$\Sigma_0=\text{diag}([10^2\ 10^2\ 4^2\ 4^2])$. The value of parameters are chosen as $a=1$, the sampling period, $T=0.2 \text{s}$, and the nominal value of the turn rate, $\Omega=3^o/\mathrm{s}$. In this problem, the uncertainty in the assumed process model is incorporated in the turn rate values.}

\begin{figure}[!t]
	\centering
	\includegraphics[width=3.2in,height=4cm]{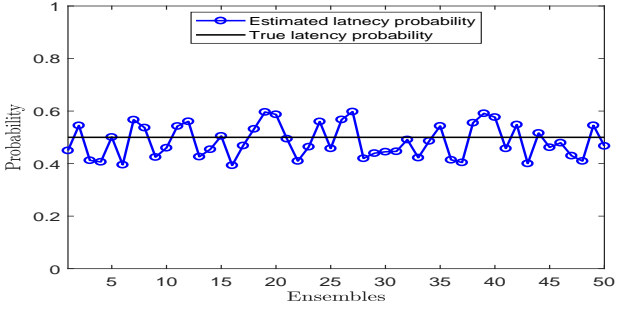}
	\caption{ Estimated latency probability.}
	\label{lat_2}
\end{figure}

The latency probability is estimated using Algorithm 1 and plotted in Fig.~\ref{lat_2} for each ensemble. The average of the estimated values of latency probability is $0.481$, whereas the true value of $\alpha$ is taken as $0.50$. The performance metrics used for this problem  to compare the different filters are RMSE and Avg-MSE in position and velocity. The RMSE in position at any time step $k$, can be defined as
{$RMSE_{pos_k}=\sqrt{\sum_{m=1}^M\left(\{(\eta_{k}-\hat{\eta}_k)^2\}_m+\{(\nu_{k}-\hat{\nu}_k)^2\}_m\right),}$} where $M$ denotes the total number of Monte Carlo runs. 500 Monte Carlo runs are used to calculate the RMSE and Avg-MSE in position and velocity. As mentioned earlier, the risk sensitive parameter, $\mu_{1,k}$ is calculated at each step, which converges at the value of $0.0013$.

\begin{figure}
	\centering
	\includegraphics[width=3.2in,height=4cm]{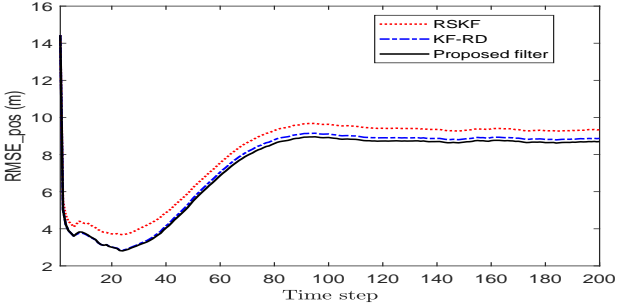}
	\caption{{ RMSE in position with $\delta\Omega=2\Omega/3$ and $\alpha=0.60$.}}
	\label{rmse_pos}
\end{figure}

\begin{figure}
	\centering
	\includegraphics[width=3.2in,height=4cm]{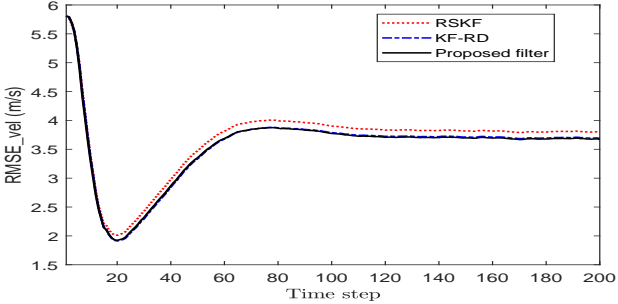}
	\caption{ {RMSE in velocity with $\delta\Omega=2\Omega/3$ and $\alpha=0.60$.}}
	\label{rmse_vel}
\end{figure}

The RMSE in position and velocity are plotted in Figs. \ref{rmse_pos} and \ref{rmse_vel}, respectively, and the uncertainty in turn rate is considered, $\delta \Omega=2\Omega/3$, when the unknown latency probability is taken as $\alpha=0.6$. From the plots, it can be observed that the proposed filter is better than the other two existing filters, and the KF-RD performs better than the RSKF. {It should also be noted that the nature of the plots are similar because all implemented filters are variants of the Kalman filter and the system is linear. Once the filters settle at some values, they remain almost there as the system parameters considered in the simulation are time invariant.} 
%

\textcolor{blue}{\begin{table}
\begin{center}
	\caption{{The Avg-MSE of different filters for various $\alpha$ and $\delta\Omega$.}}\label{Tab:2}
	\begin{tabular}{|p{0.2cm}|p{1.2cm}||p{0.8cm}|p{0.8cm}|p{0.81cm}|p{0.8cm}|p{0.8cm}|}
		\hline
		\centering \multirow{2}{*}{$\delta\Omega$}&\multirow{2}{*}{}& \multicolumn{5}{c|}{$\alpha$}\\[1ex]
		\cline{3-7}\centering
		\centering&&$0$&$0.20$&$0.40$&$0.60$&$0.80$\\[1ex]
		\hline\hline
		\centering	\multirow{3}{*}{$0$}&RSKF&3.60 &5.19 &7.84 &11.48 &16.35  \\[1ex]
		\cline{2-7}
		&KF-RD&3.59 &4.50 &5.02 &5.13 &4.92 \\[1ex]
		\cline{2-7}
		&Proposed filter &3.60 &4.52 &5.05 &5.15 &4.94\\[2ex]
		\hline
		\multirow{3}{*}{$\frac{\Omega}{3}$}&RSKF&15.70 &18.16 &22.04 &27.03 &32.68  \\[1ex]
		\cline{2-7}
		&KF-RD&16.22 &17.37 &18.51 &19.35 &19.32   \\[1ex]
		\cline{2-7}
		&Proposed filter &15.70 &16.85 &17.97 &18.79 &18.74  \\
		\hline
		\multirow{3}{*}{$\frac{\Omega}{2}$}&RSKF&30.79 &34.48 &39.15 &44.52 &51.61  \\[1ex]
		\cline{2-7}
		&KF-RD&31.96 &34.07 &35.68 &36.41 &37.36   \\[1ex]
		\cline{2-7}
		&Proposed filter &30.79 &32.86 &34.40 &35.10 &35.99 \\[2ex]
		\hline
		\multirow{3}{*}{$\frac{2\Omega}{3}$}&RSKF&51.99 &56.38 &62.05 &68.99 &76.93  \\[1ex]
		\cline{2-7}
		&KF-RD&54.12 &56.57 &58.96 &61.04 &62.66   \\[1ex]
		\cline{2-7}
		&Proposed filter &51.99 &54.38 &56.69 &58.67 &60.18  \\[2ex]
		\hline
	\end{tabular}
\end{center}
\end{table}}
{From Table~\ref{Tab:2}, where a parametric study is presented, it can be observed that the proposed filter performs comparable to the Kalman filter under no uncertainty in turn rate and without any random delays in measurements, whereas it performs better than the KF-RD and RSKF in presence of the same. It can also be noted that the Avg-MSE values of the proposed filter and the RSKF are equal in absence of the random delay ($\alpha_k=0$) in measurements.} 
\section{Conclusion}
This paper has presented a general framework of the Bayesian estimation for a system with a modeling uncertainty and random delays in measurements. A closed form solution for a linear Gaussian system is reached by following the derived general framework. Since the probability related to the random delays may not be known to the estimator in practice, a method based on maximizing the likelihood of the received measurement is illustrated to estimate the latency probability.  Further, using the uniformly complete observability and controllability criteria, the impact of random delay in measurements on the stability of the proposed risk sensitive estimator is also studied. The simulation results confirm that the proposed filter performs comparable to the Kalman filter under the nominal conditions and it is superior to the RSKF and the KF-RD when the system has model uncertainty and one step random delay in measurements. In one sentence, we conclude that the proposed filter would be an appropriate choice when the underlying system is likely to have the simultaneous presence of the model uncertainty and random delays.
%
\begin{ack}                               
The authors are grateful to Prof. Thia Kirubarajan, Department of Electrical and Computer Engineering, McMaster University, Canada, for his useful suggestions in carrying out this work. {The authors are also thankful to the anonymous reviewers for their valuable suggestions which helped to uplift the quality of the paper significantly.}  
\end{ack}
\appendix
\section{Simplification of Eq.~(\ref{joint_y})}\label{A}         
Consider the following matrix equivalence (see Appendix~A of \cite{wang2013gaussian}): 
\begin{equation}\label{C1}\
	\begin{split}
		{\begin{bmatrix} \Sigma_k^{yy} & \Sigma_k^{yx}\\\Sigma_k^{xy}&\Sigma_k^{xx}\end{bmatrix}}^{-1}
		=\begin{bmatrix}V_k^{yy} &V_k^{yx}\\V_k^{xy} & V_k^{xx}\end{bmatrix},
	\end{split}
\end{equation}
where\begin{eqnarray*}
	\begin{split}
		V_k^{yy}&=(\Sigma_k^{yy}-\Sigma_k^{yx}(\Sigma_k^{xx})^{-1}\Sigma_k^{xy})^{-1},\\ V_k^{yx}&=-(\Sigma_k^{yy}-\Sigma_k^{yx}(\Sigma_k^{xx})^{-1}\Sigma_k^{xy})^{-1}\Sigma_k^{yx}(\Sigma_k^{xx})^{-1},\\
		V_k^{xy}&= -(\Sigma_k^{xx})^{-1}\Sigma_k^{xy}(\Sigma_k^{yy}-\Sigma_k^{yx}(\Sigma_k^{yy})^{-1}\Sigma_k^{xy})^{-1},\\
		V_k^{xx}&= (\Sigma_k^{xx}-\Sigma_k^{xy}(\Sigma_k^{yy})^{-1}\Sigma_k^{yx})^{-1},
	\end{split}
\end{eqnarray*}
and all the matrices are assumed to be invertible.  
Using (\ref{C1}), we expand (\ref{joint_y}) into the following:
\begin{equation}\label{C2}
	\begin{split}
		&p(x_k,y_k|\mathcal{I}_{k-1},e_{k-1|k-1})\\&=\exp-\big[\dfrac{1}{2}(y_k-\mathbb{E}[y_k|\mathcal{I}_{k-1},e_{k-1|k-1}])^\top V_k^{yy}(y_k-\\&\mathbb{E}[y_k|\mathcal{I}_{k-1},e_{k-1|k-1}]) +\dfrac{1}{2}(y_k-\mathbb{E}[y_k|\mathcal{I}_{k-1},e_{k-1|k-1}])^\top\\&\times V_k^{yx}(x_k-\hat{x}_{k|k-1})+\dfrac{1}{2}(x_k-\hat{x}_{k|k-1})^\top V_k^{xy}\times\\&(y_k-C_k\hat{x}_{k|k-1})+\dfrac{1}{2}(x_k-\hat{x}_{k|k-1})^\top V_k^{xx}(x_k-\hat{x}_{k|k-1})\big].
	\end{split}
\end{equation}
For an arbitrary symmetric matrix, $F$, and the vectors, $g$, $D$, and $c$, consider the following identity \cite{petersen2008matrix}:
\begin{eqnarray}\label{C3}
	\begin{split}
		\dfrac{1}{2}g^\top Fg+g^\top D+c=&\dfrac{1}{2}(g+F^{-1}D)^\top F(g+F^{-1}D)\\&+c-\dfrac{1}{2}D^\top F^{-1}D.
	\end{split}
\end{eqnarray} 
Now, take $g=x_k-\hat{x}_{k|k-1},\ F=V_k^{xx}, \ D=V_k^{xy}(y_k-\mathbb{E}[y_k|\mathcal{I}_{k-1},e_{k-1|k-1}]), \ c=\dfrac{1}{2}(y_k-\mathbb{E}[y_k|\mathcal{I}_{k-1},e_{k-1|k-1}])^\top\\ \times V_k^{yy}(y_k-\mathbb{E}[y_k|\mathcal{I}_{k-1},e_{k-1|k-1}])$, and process (\ref{C2}) in accordance with (\ref{C3}).  After absorbing the terms which do not contain $x_k$ into the normalizing constant, $\gamma_{k|k}$, and using equivalence of matrices from (\ref{C1}),  we obtain (\ref{cond2}). Note that in the final expressions, $\Sigma_k^{yx}$ is replaced with $\Sigma_k^{xy^\top}$. 
{\section{Proof of Theorem~\ref{stabality}}\label{stability_proof}
%
	The least square estimate of $x_k$, based on recent $l$ observations after ignoring the process noise and considering the nominal process model, can be given as (see Lemma~7.1 of \cite{jazwinski2007stochastic})
	\begin{equation}\label{x_least}
		\bar{x}_{k|k-l:k}=O_{k,k-l}^{-1}\sum_{i=k-l}^k A_{i,k}^\top C_i^\top R_i^{-1}z_i, \quad k\geq l.
	\end{equation}
	The estimate, $\bar{x}_{k|k-l:k}$, is suboptimal \cite{jazwinski2007stochastic} and, consequently,
	\begin{equation}\label{cov_ineq}
		\mathbf{P}_{k|k} \leq \mathbb{E}[(x_k-\bar{x}_{k|k-l:k})(x_k-\bar{x}_{k|k-l:k})^\top].
	\end{equation}
	To compute the covariance in (\ref{cov_ineq}), from (\ref{sys}) and (\ref{trans_mat}), we can write
	\begin{equation}\label{x_i}
		\begin{split}
		x_i=&(A_{i,k}+\Delta A_{i,k})x_k -(A_{i,k}+\Delta A_{i,k})\\&\times\sum_{j=i}^{k-1}(A_{k,j+1}+\Delta A_{k,j+1})w_j,
		\end{split}
	\end{equation}
	and substituting the expression of $z_i$, obtained from (\ref{x_i}) and (\ref{nd_lin_meas}), into (\ref{x_least}), we can write
	\begin{equation*}
		\begin{split}
			&(x_k-\bar{x}_{k|k-l:k})=x_k-O_{k,k-l}^{-1}\sum_{i=k-l}^k A_{i,k}^\top C_i^\top R_i^{-1} C_i\\&\times(A_{i,k}+\Delta A_{i,k})x_k+O_{k,k-l}^{-1}\sum_{i=k-l}^kA_{i,k}^\top C_i^\top R_i^{-1}C_i\\&\times (A_{i,k}+\Delta A_{i,k})\sum_{j=i}^{k-1} (A_{k,j+1}+\Delta A_{k,j+1})w_j\\&
			-O_{k,k-l}^{-1}\sum_{i=k-l}^kA_{i,k}^\top C_i^\top R_i^{-1}v_i.
		\end{split}
	\end{equation*}
	Using the definitions of $\Delta O_{k,k-l}$, the above expression can be reduced as
	\begin{equation*}
		\begin{split}
			&(x_k-\bar{x}_{k|k-l:k})=-O_{k,k-l}^{-1}\Delta O_{k,k-l} x_k\\& + O_{k,k-l}^{-1}\sum_{i=k-l}^kA_{i,k}^\top C_i^\top R_i^{-1}C_i (A_{i,k}+\Delta A_{i,k})\times\\&\sum_{j=i}^{k-1} (A_{k,j+1}+\Delta A_{k,j+1})w_j
			-O_{k,k-l}^{-1}\sum_{i=k-l}^kA_{i,k}^\top C_i^\top R_i^{-1}v_i,
		\end{split}
	\end{equation*}
	and
	\begin{equation*}
		\begin{split}
			&\mathbb{E}[(x_k-\bar{x}_{k|k-l:k})(x_k-\bar{x}_{k|k-l:k})^\top]\\
			&=O_{k,k-l}^{-1}\Delta O_{k,k-l}\mathbb{E}[x_kx_k^\top]\Delta O_{k,k-l}^\top {O_{k,k-l}^{-1^\top}}\\&\quad+\mathbb{E}\Big[\Big(O_{k,k-l}^{-1}
			\sum_{i=k-l}^kA_{i,k}^\top C_i^\top R_i^{-1}C_i (A_{i,k}+\Delta A_{i,k})\\&\quad\times\sum_{j=i}^{k-1} (A_{k,j+1}+\Delta A_{k,j+1}) w_j\Big)\Big(\ast\Big)^\top\Big]\\&\quad +\mathbb{E}\Big[\Big(O_{k,k-l}^{-1}\sum_{i=k-l}^kA_{i,k}^\top C_i^\top R_i^{-1}v_i\Big)\Big(*\Big)^\top\Big]\\
			&\leq O_{k,k-l}^{-1}\Delta O_{k,k-l}\mathbb{E}[x_kx_k^\top]\Delta O_{k,k-l}^\top {O_{k,k-l}^{-1^\top}}\\&\quad+\mathbb{E}\Big[\Big(O_{k,k-l}^{-1}
			\sum_{i=k-l}^kA_{i,k}^\top C_i^\top R_i^{-1}C_i (A_{i,k}+\Delta A_{i,k})\\&\quad\times\sum_{j=k-l}^{k-1} (A_{k,j+1}+\Delta A_{k,j+1}) w_j\Big)\Big(\ast\Big)^\top\Big]\\&\quad +\mathbb{E}\Big[\Big(O_{k,k-l}^{-1}\sum_{i=k-l}^kA_{i,k}^\top C_i^\top R_i^{-1}v_i\Big)\Big(\ast\Big)^\top\Big],
		\end{split}
	\end{equation*}
	where $(\ast)$ represents the same terms that are given in the parenthesis left to it.  Now, using the relation $\mathbb{E}[x_kx_k^\top]=\mathbf{P}_{k|k}+\hat{x}_{k|k}\hat{x}_{k|k}^\top$ and (\ref{cov_ineq}), and rearranging the terms, we have
	\begin{equation}\label{before_norm}
		\begin{split}
			&\mathbf{P}_{k|k}-O_{k,k-l}^{-1}\Delta O_{k,k-l}\mathbf{P}_{k|k}\Delta O_{k,k-l}^\top {O_{k,k-l}^{-1^\top}}\\&\quad\leq 
			O_{k,k-l}^{-1}\Delta O_{k,k-l}\hat{x}_{k|k}\hat{x}_{k|k}^\top\Delta O_{k,k-l}^\top {O_{k,k-l}^{-1^\top}}\\
			&\qquad+(\mathbf{I}+O_{k,k-l}^{-1}\Delta O_{k,k-l})\mathcal{C}_{k,k-l}(\mathbf{I}+O_{k,k-l}^{-1}\Delta O_{k,k-l})^\top\\&\qquad+ O_{k,k-l}^{-1}.
		\end{split}
	\end{equation}
	Taking the Euclidean norm $(||.||)$ on the both sides of (\ref{before_norm})  and writing the relation for the left hand side of it, we have
	\begin{eqnarray}\label{ineq_rel}
		\begin{split}
			&||\mathbf{P}_{k|k}||-||O_{k,k-l}^{-1}\Delta O_{k,k-l}\mathbf{P}_{k|k}\Delta O_{k,k-l}^\top {O_{k,k-l}^{-1^\top}}||\\&\leq 	||\mathbf{P}_{k|k}-O_{k,k-l}^{-1}\Delta O_{k,k-l}\mathbf{P}_{k|k}\Delta O_{k,k-l}^\top {O_{k,k-l}^{-1^\top}}||.
		\end{split}
	\end{eqnarray}
	Again, using the relation 
	\begin{equation*}
		\begin{split}
			&||O_{k,k-l}^{-1}\Delta O_{k,k-l}\mathbf{P}_{k|k}\Delta O_{k,k-l}^\top {O_{k,k-l}^{-1^\top}}||\\&\leq 	||O_{k,k-l}^{-1}\Delta O_{k,k-l}||\ ||\mathbf{P}_{k|k}||\ ||\Delta O_{k,k-l}^\top {O_{k,k-l}^{-1^\top}}||
		\end{split}
	\end{equation*}
	in (\ref{ineq_rel}) and substituting it in (\ref{before_norm}) with the norm, we can write
	\begin{equation}\label{final_ineq}
		\begin{split}
			&||\mathbf{P}_{k|k}|| \leq ||O_{k,k-l}^{-1}\Delta O_{k,k-l}\hat{x}_{k|k}\hat{x}_{k|k}^\top\Delta O_{k,k-l}^\top {O_{k,k-l}^{-1^\top}}\\
			&\quad+(\mathbf{I}+O_{k,k-l}^{-1}\Delta O_{k,k-l})\mathcal{C}_{k,k-l}(\mathbf{I}+O_{k,k-l}^{-1}\Delta O_{k,k-l})^\top\\&\quad+ O_{k,k-l}^{-1}||/(1-||O_{k,k-l}^{-1}\Delta O_{k,k-l}||^2).
		\end{split}
	\end{equation}
	Recalling the assumption that the states are bounded and using the Eq.~(\ref{obs}), it is evident that if the uncertainty in process model is finite and follows the condition, $-\mathbf{I}<O_{k,k-l}^{-1}\Delta O_{k,k-l}<\mathbf{I}$, $\mathbf{P}_{k|k}$ in (\ref{final_ineq}) is bounded for all $k\geq l$.}
\bibliographystyle{plain}        
\bibliography{rs_1d}           
%
%
\end{document}